\documentstyle{article}

\begin{document}
\centerline{ARE THERE DYNAMICAL LAWS?}
\smallskip
\centerline{J. Anandan}
\centerline{Department of Physics and Astronomy}
\centerline{University of 
South
Carolina}
\centerline{Columbia, SC 29208, USA.}
\centerline{E-mail: jeeva@sc.edu}
\centerline{and}
\centerline{Foundations of Physics}
\centerline{Buys Ballotlaboratorium}
\centerline{University of Utrecht}
\centerline{Princetonplein 5}
\centerline{3584 CC Utrecht, The Netherlands}

\begin{abstract}

The nature of a physical law is examined, and it is suggested that 
there may not be any fundamental 
dynamical laws. This explains the intrinsic indeterminism of 
quantum theory. The probabilities for transition from a given 
initial state to a final state then depends on the quantum 
geometry that is determined by
symmetries, which may exist as relations between states
in the absence of dynamical laws. 
This enables the experimentally well confirmed quantum
probabilities to be derived from the geometry of Hilbert space, 
and gives rise to effective probabilistic laws. An arrow of time which is 
consistent with the one given by the second law of 
thermodynamics, regarded as an effective law, is obtained. Symmetries are used 
as the basis for a new proposed paradigm of physics. This gives rise naturally 
to the gravitational and gauge fields from the symmetry group of the standard 
model, and a general procedure for obtaining interactions from any symmetry 
group.
\end{abstract}

\bigskip
Aug. 23, 98. Revised July, 31, 99. Quant-ph/9808045.

\newpage
\section{The Paradigm of Laws}

At the end of the millennium we are faced with, what may be the most difficult 
intellectual challenge known to humankind, the problem of constructing a quantum 
theory of gravity that would unify all the interactions. The great difficulty of 
this problem is underlined by the fact that it has remained unsolved for more 
than seven decades. There is another profoundly difficult problem at the heart 
of quantum theory, which has also remained unsolved for the same period of time, 
namely the quantum measurement problem. This article will suggest that the 
present insolubility of both problems may have the same source, namely the 
paradigm of physics that has been used since the creation of physics about four 
centuries ago. I shall consider now, briefly, the origin of this paradigm, and 
argue that it is based on the metaphysical assumption of the existence of 
dynamical laws, which may be discarded.

Prior to the origin of physics, `explanations' of the world were 
often
attempts to describe the boundary conditions, such as how the 
world came
into being. But some natural philosophers recognized that much 
more
progress could be made in `understanding' the world if we gave up 
trying
to explain boundary conditions and instead described the universal
regularities that seem to occur in contingent phenomena.  They 
tried
to predict what would happen if certain arbitrary initial conditions 
were
specified, instead of explaining these initial conditions 
(see, for example, Wigner, 1967, p. 40).
The tool for making this prediction is called a dynamical law of 
physics. More
precisely, a dynamical law of physics, which I shall call a {\it law } for 
simplicity in this paper, may be
defined as the ability to describe the initial state of a physical 
system
from which the final state can be predicted, deterministically or
probabilistically, using the nature of the system and its interaction 
with
its environment.

This profound realization led to the origin of physics. The 
combination of contingencies, reflected in the boundary conditions 
or 
initial conditions, and universal laws that are 
independent of 
contingencies has been the paradigm of physics for four centuries. 
This will be called the paradigm of laws. It is a curious fact, however, 
that the laws we use are always 
associated with symmetries. This follows from the fact that
the experiments allowed by these laws are {\it 
reproducible} 
in different places and times, with
different orientations etc. (Wigner, 1967, p. 29). Moreover, the symmetries of 
the laws may be thought of as
determining the physical geometry (Anandan, 1980a). Conversely, we 
may start with the 
geometry and conclude that the laws are constrained 
by the 
symmetries of the geometry.

Now, a law of physics is a strange type of
necessity that should be confirmable or refutable (i.e. testable) in order for 
it to 
carry information about the world. So it cannot be a logical 
necessity, 
which is tautological and therefore is not refutable. Belief in 
such a
metaphysical necessity of laws smacks of the belief in a 
supernatural agency that is always regulating natural phenomena. 

I shall therefore explore the view that 
there is no such metaphysical necessity, which implies that there
are no fundamental laws of physics. Pierce (1891), Wheeler (1980, 1985, 1990, 
1994), and Smolin (1997) have suggested that the laws of physics are mutable. 
But here I shall question even the very existence of laws, and propose a new 
paradigm in which `dynamical 
laws' will be replaced by symmetries as the fundamental 
relational structures in the world. Van Fraassen (1989) has made an empiricist 
critique of any kind of structural realism such as laws or symmetries, but 
regards symmetries as a better clue to theorizing than laws. While I also do not 
regard laws as real, I shall however regard symmetries as real to be associated 
with the quantum geometry which gives rise to all the observed interactions, 
analogous to how space-time geometry gave rise to gravitation in classical 
general relativity.

The non existence of laws imply that there can be neither deterministic laws 
(sections 2,3,4) nor fundamental proabilistic laws (section 5). The 
probability rule of quantum theory will be deduced, in section 5, from the 
quantum geometry and the associated symmetries, and 
therefore will not have the status of a fundamental law. This approach naturally 
gives an arrow of time (section 3). In section 6, the modular energy-momentum of 
Aharonov et al (1969) will be used characterize the four fundamental 
interactions in terms of the symmetry group of the standard model. This is then 
generalized to a new principle which gives the interactions corresponding to the 
symmetry group of any physical theory. In section 7, I shall generalize the 
celebrated Erlanger program of Klein (1872) for constructing and defining a 
geometry as the set of properties that are invariant under a symmetry group 
acting on a set of points, by giving up the requirement that there should be 
such a set in quantum theory. The advantages of the present approach over 
	other interpretations of quantum theory are pointed out in section 8. The 
relevance of this approach to the construction of quantum gravity is discussed 
in the concluding section 9.

\section{Processes and Phenomena}

Throughout this article, the term `state' refers to a state in which an
individual system has been observed to be. It is not a state that 
has
been constructed statistically by measurements on an ensemble of 
identical
systems, as it is in the Copenhagen interpretation of quantum theory. The 
justification for this viewpoint and a precise mathematical description of the 
state, based on possible observations of an individual system, will be given in 
section 4.

The absence of deterministic laws implies that
there need not be a unique time evolution for any physical 
system. 
This will now be stated as
an assumption, called $A_1$.

\medskip
\noindent{$A_1$\it: Physical systems that start from the same initial state 
need not end up in 
the same final state under the same experimental conditions. }
\medskip

Define therefore a {\it phenomenon} to be a sequence of pairs
$\{(\alpha,\beta_1),(\alpha,\beta_2),...(\alpha,\beta_n),....\}$ in an 
experiment such that $\alpha$ is the common initial state of identical systems 
and 
$\beta_i$ are their final states under 
the 
same experimental conditions. The $\beta_i$s need not all be 
distinct. Each pair $(\alpha,\beta_i)$, regarded as an element of a phenomenon, 
will be called a trial. Distinct trials will be called {\it processes}. In the 
same experiment and for the same system, there can be trials with an initial 
state $\alpha'$ that is distinct from $\alpha$, but these would then belong to a 
different phenomenon. (This remark will be relevant to the thought experiment 
that will 
be considered in the next section.)

We know that the 
observation of quantum phenomena is consistent with $A_1$, 
although no attention was paid  to 
quantum physics in my critique of the metaphysical assumption that there are 
laws that led to $A_1$. An example is a 
Stern-Gerlach experiment with neutrons in 
which the inhomogenity of the 
magnetic field is in the $z-$direction, and each incoming neutron 
is observed as it enters the apparatus to have its
spin in the $x-$direction, which corresponds to the initial state 
$\alpha$. The neutrons then end up in the two possible final states, $\beta_1$ 
and $\beta_2$, at the two spots on the screen, corresponding to the spin up 
state and the spin down states 
with respect to the $z-$direction.

By the {\it probability} of a process $(\alpha,\beta)$ is meant the relative 
frequency of this pair in a phenomenon as the number of trials tends to infinity 
under the same experimental conditions. Since all trials in the phenomenon have 
the same initial state, this is the conditional probability of observing the 
system in $\beta$ if it was previously observed in state $\alpha$.  These 
probabilities are well defined in the sense that the probability of a process is 
the same for phenomena that have the same initial state as the given process 
under the same experimental conditions. A {\it symmetry} is a 
transformation on the set of states with the property that the probabilities of 
processes are 
invariant under this transformation. Later I shall associate symmetries {\it a 
priori} with the quantum geometry which will then {\it 
explain} why
processes have well defined probabilities and why these probabilities are 
invariant under symmetries. 

The apparent existence of deterministic laws in classical physics is a mystery, 
given $A_1$. But 
this mystery is
removed by the realization that the phenomena that make up 
classical
physics are limiting cases of quantum phenomena that correspond 
to the
probability of some processes being much greater than others. 
This gives
the illusion of deterministic laws in classical physics.  E.g. suppose 
a
baseball or cricket ball is thrown and the positions of its
center of mass are observed to some uncertainty in a series of
measurements. These positions are very likely to be on an approximate
parabola. But there is no law compelling them to lie on a parabola 
as was
mistakenly assumed in classical physics. 

\section{ The Arrow of Time}

In general, a phenomenon is not time-reversible. This may be 
illustrated 
by an example due to Penrose (1989, p. 357). He 
considered 
photons emitted from a lamp in a state, denoted here by 
$\alpha_1$, 
that is aimed at a detector. Between the lamp and the detector 
there 
is a half-reflecting mirror inclined at an angle, say $45^0$, to the 
line 
connecting the lamp to the detector. The probability of 
transmission 
or reflection of the photon by the glass is $1\over 2$ for incidence 
on
either side. Then 
as the number of photons emitted in state $\alpha_1$ tends 
to infinity, half of them are detected in state $\beta_1$, say, by 
the 
detector, while the other half, reflected by the mirror, strike a 
wall where they are observed 
in a 
state $\beta_2$, say. So this experiment gives a phenomenon 
such as $P\equiv \{ (\alpha_1,\beta_1), 
(\alpha_1,\beta_2),(\alpha_1,\beta_2),(\alpha_1,\beta_1),....\}$. 
This 
is consistent with $A_1$ above. Now Penrose points out that 
while the probability of $\beta_1$ given $\alpha_1$ is $1\over 
2$, 
the 
probability of $\alpha_1$ given $\beta_1$ is $1$. I.e. the 
probability of the process $(\alpha_1,\beta_1)$ forwards in time 
is 
not the same as the probability of this process backwards in time. 
More generally, it follows from the above definition of a 
phenomenon 
that the probabilities forwards and backwards in time of any 
process 
in it are not the same unless the phenomeonon is such 
that all the final states $\beta_i$ are the same.

The time-reversal of the Penrose phenomenon $P$ above is the 
sequence 
$P^*\equiv 
\{(\beta_1^*,\alpha_1^*), 
(\beta_2^*,\alpha_1^*),(\beta_2^*,\alpha_1^*),(\beta_1^*,\alpha_1^*
),...
\}$, where the * denotes the time-reversal of the unstarred state 
(in 
the present case * changes absorption to emission and vice versa).  
The subsequence of $P^*$ that consists of pairs with the same first component 
$\beta_1^*$ is $P_1^*\equiv 
\{ (\beta_1^*,\alpha_1^*), (\beta_1^*,\alpha_1^*) ,(\beta_1^*,\alpha_1^*
),...\}$. If $P_1^*$ were a phenomenon then the probability of the process 
$(\beta_1^*,\alpha_1^*)$ would be $1$. But $P_1^*$ cannot take place in the 
forward time direction if we 
assume 
rotational symmetry of $P$. To see this,
simply rotate the apparatus about the axis  perpendicular to the 
plane of the apparatus through the middle of the 
half-reflecting mirror by $180^0$ so
that the positions of the lamp and the detector are interchanged. 
The 
assumption now that $P$ is invariant under this rotation implies 
that 
we could have 
the phenomenon $P_R$ obtained from $P$ by the above mentioned rotation, i.e.
$P_R=\{(\beta_1^*,\alpha_1^*),(\beta_1^*,\alpha_2^*),....\}$, 
with each of the two processes having probability $1/2$, where 
$\alpha_2^*$ is the state of the photon as it is being absorbed, 
after reflection by the mirror, 
in the 
wall opposite to the wall on which $\beta_2$ is absorbed. 
So $\beta_1^*$ does not always end up as $\alpha_1*$ as required 
by $P_1^*$.
Therefore, $P_1^*$ is not allowed in the forward time direction, 
Hence,
by observing $P$, we can determine the direction of time. Another way of saying 
this is that if a movie of this experiment is taken, then when the movie is 
viewed we can say whether the movie is running forwards or backwards, which 
therefore gives an arrow of time.

It may be argued that a photon may be emitted, albeit with very low probability, 
in the state  $\alpha_2$, that is the time reversal of $\alpha_2^*$, from the 
corresponding wall, which may then end up as $\beta_1$ or $\beta_2$ with equal 
probability. If this occurs then $P_1^*$ should be modified to include the 
process $(\beta_1^*,\alpha_2^*)$. But this process has a very low probability, 
close to $0$, in $P_1^*$, whereas in $P_R$ it has probability $1/2$. So, 
rotational symmetry again prevents $P_1^*$ being a phenomenon in the forward 
time direction. This consideration shows that if all possible initial and final 
states are included, the violation of time reversal symmetry is due to the non 
invariance of probabilities under time reversal.

Each 
phenomenon distributes systems in the same or similar original 
states to different states with higher entropy, in 
general. And the time-reversal of a phenomenon which would 
lead to 
decrease in entropy, in general, cannot occur. Hence, the above 
arrow 
of time is the same as the arrow of time 
given by the second law of thermodynamics.
Here the second law of thermodynamics is regarded as an 
effective law
and not a fundamental law. By this I mean that the 
second law of thermodynamics is a statistical statement\footnote{To quote 
Wheeler (1980), ``Ask any molecule what it thinks about the second law of 
thermodynamics and it will laugh at the question. All the same the molecules, 
collectively, uphold the second law.''}  about the 
outcomes in a large number of experiments, none of which is 
governed by a law within the framework of quantum 
theory. The inherent probabilities in the individual processes, it 
will be argued later, in section 5, are the outcomes of the quantum geometry 
that is determined by symmetries. So, even this statistical statement is {\it 
not} a fundamental law.

This appears to resolve at least partially the well known paradox 
which arises 
in the paradigm of laws due to the fact that the 
microscopic laws have time-reversal symmetry (ignoring weak 
interactions which are not relevant to this problem which arises 
even 
in the absence of weak interactions), and yet there is an 
observable 
direction of time. In the present approach, because there are no 
laws, 
this problem does not arise.

The time-irreversibility mentioned above should be distinguished 
from the `irreversible' amplification which occurs when a 
measurement is made on the system, which Penrose (1989)
calls the $R$-process (stands for the `reduction of the state 
vector'). 
The latter irreversibility is associated only with the measurement 
or 
observation of {\it each} of the initial and final states of an 
individual system, whereas the former irreversibility 
is associated with {\it both} the initial and final states in a 
phenomenon, which is given physical meaning by means of an 
ensemble. 

\section{Meaning of States And Processes}

The question naturally arises as to what exactly are the initial 
and 
final states that were referred to in the previous two sections. An operational 
meaning may be given to states from the fact that it is 
possible to observe the extended wave function of a {\it single} 
system by means of protective measurements, which were 
introduced by Aharonov, Anandan and Vaidman (1993). For example if the system is 
in a non degenerate eigenstate of the Hamiltonian, then an adiabatic 
measurement on it, which does not lead to entanglement between the system and 
the apparatus, would be a protective measurement.
This protective observation therefore shows that an extended 
wave 
function is real. But it is not necessary to assume the Hilbert space 
structure, including wave functions, in doing protective 
measurements. One may begin with the set of physical states, or 
simply states, of a physical system as an abstract set without any 
structure on it, as in the present approach.
Then, as shown by Anandan (1993a), from the numbers that an experimentalist 
could in 
principle 
obtain from protective measurements on these states for various 
observables, the entire 
Hilbert space structure may be constructed; the states are then 
rays in this Hilbert space and the numbers obtained by the 
experimentalist are of the form $<\psi|A|\psi>$, where $|\psi>$ is a 
normalized vector in the ray representing the state and $A$ is a 
Hermitian operator now representing the observable that was 
protectively measured. In the latter approach, due to the absence of the Hilbert 
space structure at the beginning, it is not possible to determine the state by 
calculation prior to the protective measurements, as one could do if there were 
a Hilbert space structure and one knows the protecting Hamiltonian.

Moreover the following purely empirical criterion may be used to 
ensure 
and recognize that a given experiment is a protective 
measurement and
is not the usual measurement: If we make repeated 
measurements on a single system
of any pair of 
{\it arbitrary} observables $A,B$ which are alternated indefinitely 
(i.e. measure $A,B,A,B,...$), 
then only if the measurements are protective are the same values 
{\it always}
obtained for $A$ and $B$ (which are $<\psi|A|\psi>$ 
and $<\psi|B|\psi>$ respectively). In order for this to be the case 
for
the usual measurements it would be necessary for $A$ and $B$ to 
commute. This empirical criterion does not require any knowledge of the 
Hamiltonian. Therefore, this criterion and the observation at the end of the 
previous paragraph refute the criticism that one could obtain the wave function 
by calculation if the Hamiltonian is known and therefore no new information is 
obtained by protective measurements. Hence, 
protective observations of states have the same epistemological status as the 
usual measurements, and have the advantage over the latter of establishing the 
ontology of the states.

I shall therefore take the possible initial or final states of the 
system 
(i.e. the $\alpha$s and $\beta$s in the sections 2 and 3) to be any 
rays 
in the Hilbert space of the system. This is because the theorem in the first 
paragraph of this section says that the Hilbert space may be constructed from 
protective measurements and, conversely, any ray in the Hilbert space 
may be protectively observed, in principle. For example, suppose an electron in 
an atom is in an excited state of energy, and decays to the ground state. Then 
both its initial and final states may, in principle, be protectively observed. 
So in the present approach 
there 
is no `preferred basis'. 

By two states being orthogonal I mean that 
any pair of vectors belonging to the rays representing these states 
is 
orthogonal. The final states that are not the same (i.e.  distinct) are 
mutually exclusive and are 
assumed to be orthogonal, which is contained in the second 
assumption:

\medskip
\noindent {$A_2$. \it Any two distinct possible final states of a phenomenon are 
orthogonal. }
\medskip

The essential reason for the time irreversablility of a phenomenon that has 
distinct final 
states, mentioned in section 2, can now be understood in the usual 
quantum theory as due to the unitarity of time evolution. This has been 
called the $U-$process by Penrose (1989). This implies that 
orthogonal states must necessarily evolve into orthogonal states. 
Hence it is not possible for identical systems 
beginning from different initial states to evolve always to the 
same final state. Therefore, if measurements are made to determine the final 
state for systems having two possible initial states we cannot always obtain the 
same final state. But if a phenomenon such as $P$ in section 2 is time reversed, 
then systems in distinct initial states will always go to the same final state, 
which violates unitarity. So, the time-reversal of a phenomenon 
cannot be realized, except in the special case when the experimental 
conditions are such that only one final state is possible. Unitary evolution 
preserves information. But when there is one initial state and more than one 
final states for an ensemble of systems, as it is for a phenomenon in general, 
information is lost. This loss of information is equivalent to an increase in 
entropy as mentioned in the previous section.

Within the paradigm of laws there is now the following 
paradox: Suppose a particle's state is protectively observed 
assigning 
to it an extended wave function $\psi_1$, up to an arbitrary phase 
factor and gauge transformations if the particle is charged. This 
direct observation of the wave function implies that it is real. 
Suppose now that a usual measurement is made on the system 
and 
the system is found to have a localized wave function $\psi_2$, 
which is equally real. How do we account for this sudden 
transition 
from $\psi_1$ to $\psi_2$?

One may try to explain this as being due 
to a dynamical `collapse', i.e. a transition which is governed by 
some 
law (Pearle, 1986; Ghiradi, Rimini and Weber, 1986; 
Pearle, 1989; Diosi, 1989; Ghiradi, Grassi and Rimini, 1990; 
Penrose, 1996). But dynamical collapse models generally have 
the 
following three problems: a) They violate energy-momentum 
conservation. Even if this happens so rarely that it is unnoticed, 
it is unnatural nevertheless from the present point of view which 
gives great importance to symmetry principles and the associated 
conserved quantities. b) If the wave function is charged we would 
expect it to radiate as it `collapses' from $\psi_1$ to $\psi_2$. c) It 
has not been possible, as far as I know, to construct a satisfactory 
Lorentz covariant model of the `collapse'. But these problems and 
the above 
paradox disappear if 
we give up the notion of a law for an individual system.
 
The above considerations show the problems which arise if we 
require that each individual system obeys the $U$-process. For the 
individual system what is observed is the $R$-process and not the 
$U$-process. (In protective measurement the $U$-process is 
observed 
for the individual system but this is in the trivial case where the 
$U$ 
process does not change the state.) Hence if there is a conflict 
between the two processes 
then we should give up the $U$-process for the individual system. 
The 
wave function being associated with a single system by 
protective measurements and the observed $R$ 
process that the wave function of this system undergoes suggest 
that the transition from the initial to the final wave function of the 
individual system is 
not 
governed by any law. 

On the other hand, for an ensemble of particles which 
begin with the same initial state, the $U$-process can be observed. 
E.g. in the Stern-Gerlach experiment mentioned in section 2, the 
superposition of the two wave packets that are obtained from the 
$U$-process can be observed for the entire ensemble of neutrons 
by 
the two spots they form on the screen. When an individual 
neutron 
undergoes the $R$-process into one of the spots, it is not necessary 
to 
`collapse' the wave function if it represents the entire ensemble. 
Alternatively, we can observe the $U$-process by measuring an 
Observable, one of whose eigenstates is the state obtained from the 
initial state by the $U$-process. Then all the systems in the 
ensemble 
which had the same initial state will be found in the same final 
state 
that is obtained from the $U$-process. But again, to verify this we 
need to make measurements on an ensemble of a large number of 
copies of the system.
Hence I shall associate the $U$-process with an ensemble of 
systems 
with the same initial state and the $R$-process with the individual system.

\section{Probabilities of Processes}

Assumption $A_1$ implies that 
there are no deterministic laws. This still leaves 
open the
possibility of there being probabilistic laws. I shall 
now argue even against probabilistic laws as fundamental laws. To 
do so, 
it
is necessary to derive the experimentally well confirmed
probabilities
that were postulated
in quantum theory from something more fundamental, which I 
shall do now.
The answer proposed here to Einstein's famous 
question as to
why God plays dice with the universe is that an individual system 
is forced to obey only logical necessity and therefore it cannot 
obey any dynamical law 
that is not a logical necessity. 
There can, nevertheless, be well defined probabilities as a 
consequence of
symmetries.

Consider, as an example, the tossing of a coin. The 
equal probabilities of heads and tails are due to the symmetry 
between the two sides, and are independent of the particular 
law which governs the falling of the coin. This therefore suggests 
(but does not prove) that 
the equal probabilities may exist as a consequence of the 
symmetry 
alone and may be independent of whether there is a law. A 
classical 
coin obeys (apparent) dynamical laws and  the indeterminacy 
of outcomes is due to our ignorance of the initial conditions while 
the 
equal probabilites are due to the symmetry of the coin and of the 
possible initial conditions with respect to the possible final 
outcomes. But for a quantum system, due to the absence of 
law in the present approach, there is an intrinsic indeterminacy. 
Nevertheless, there are well defined probabilities which may be 
regarded as due to the 
symmetries of the initial state with respect to the possible final 
states, as I shall show now.

Consider first the Stern-Gerlach experiment, mentioned in section 2. The 
initial wave function $\psi$ of the 
neutron is an equal superposition of the two normalized wave 
packets 
$\psi_{\uparrow}$ and $\psi_{\downarrow}$ that end up in the 
two 
spots with probability $1/2$, i.e. $\psi ={1\over \sqrt{2}} 
\psi_{\uparrow}+{1\over \sqrt{2}}\psi_{\downarrow}$. Consider 
the 
two 
dimensional Hilbert space spanned by $\psi_{\uparrow}$ and 
$\psi_{\downarrow}$. The set of rays or the projective space 
of this Hilbert space 
is a sphere with the Fubini-Study metric (Kobayashi and Nomizu, 1969; Anandan 
and Aharonov 1990; Anandan, 1991)
being 
the usual metric on this sphere. Also, $\psi_{\uparrow}$ and 
$\psi_{\downarrow}$ correspond to opposite points of this sphere 
and 
$\psi$ is half-way between them on a 
geodesic connecting them. This is somewhat analogous to the 
symmetry of the coin discussed above. It is reasonable therefore 
to 
suppose that
$\psi$ has equal probabilities of going into $\psi_{\uparrow}$ and 
$\psi_{\downarrow}$. 

Given an arbitrary quantum system, there is a group of unitary 
and anti-unitary transformations which act on its Hilbert space. 
Physically, this group 
may be thought of as relating states that belong to the {\it same} 
physical system and preserving mutual exclusivity (represented 
by 
orthogonality) of the distinct states among the final states of a 
phenomenon as required by $A_2$. 
The  properties that are 
invariant 
under this group, which include the Fubini-Study metric (Anandan and Aharonov, 
1990)
may therefore be regarded as a geometry in 
the sense of Klein's Erlanger program (Klein, 1872). 
These
properties are of 
course also invariant under the universal symmetry group of the 
fundamental interactions which is a subgroup of this meta-
symmetry
group of  unitary and anti-unitary transformations.

Define the distance between two states to be the Fubini-Study 
length 
of the shortest geodesic joining them. The distance $s$ between 
two 
states that are represented 
by normalized state vectors $|\psi>$ and $|\psi'>$ is then given by
\begin{equation}
\cos {s\over 2}= |<\psi|\psi'>|,
\label{distance}
\end{equation}
where $s\epsilon [0,\pi]$. Here I have scaled the Fubini-Study 
metric 
so that the sphere that is the projective space of the Hilbert 
subspace 
spanned by 
$\psi$ and $\psi'$ has unit radius. Then the distance between two 
orthogonal states (opposite points on the sphere) is $\pi$. In 
the
following only the definition of distance given by (\ref{distance}) 
is 
used, and no other knowledge of the Fubini-Study metric is 
needed.

The present work, to use the usual formalism of quantum theory, corresponds to 
the Heisenberg picture in which the 
$U$-process governs the observables, while the state 
is changed only during an $R$-process. Since these time 
dependent observables are common to all 
systems which have the given Hamiltonian, this is consistent with 
the 
conclusion above that the $U$- process should be associated with 
an 
ensemble of systems while the $R$-process may be associated 
with 
the individual 
system. I now generalize the 
equiprobabilities in the above Stern-Gerlach experiment to the 
following assumption:

\medskip
\noindent {\it $A_3$. Suppose a system is in an initial
state $\alpha$, which has equal distances to possible final 
states $\beta_1$ and $\beta_2$. Then the probabilities of 
transition 
from $\alpha$ to $\beta_1$ and $\beta_2$ are equal.} 
\medskip

In the Heisenberg picture the initial and final states are on the 
same 
footing because they are all time independent. Also,  in 
$A_2$, the mutual exclusivity of the final 
states of a phenomenon was expressed by their orthogonality. This 
suggests that a state that is orthogonal to the initial state 
cannot be a possible final state. Similarly, we would expect that a 
state that is orthogonal to all possible final states must be 
orthogonal 
to the initial state. The last two statements are equivalent to the 
following assumption:

\medskip
\noindent {\it $A_4$. The state vector representing the initial state in 
a phenomenon can be expressed as a linear 
combination of the state vectors that represent all possible final 
states with non zero coefficients.}
\medskip

Probabilities of transition between an arbitrary pair of states will 
now be obtained using the `classical' or `standard' definition of
probability 
(Gnedenko, 1967; Zurek, 1998). The basic idea here is to derive an 
arbitrary 
probability by expressing it in terms of equiprobabilities of 
equally 
likely events, which is a primitive concept. In the present case, 
transitions to 
states that are equidistant from the initial state have equal 
probabilities by $A_3$. 
Consider a phenomenon in which the initial state is represented 
by 
$|\psi>$ and all possible distinct final states by $|\psi_i>, i=1,2,...N$. 
Then 
from $A_2$ and $A_4$, we can write
\begin{equation}
|\psi>=\sum_{i=1}^N c_i |\psi_i>
\label{psi}
\end{equation}
where without loss of generality $c_i$ are real and positive, and
\begin{equation}
<\psi_i|\psi_j>=\delta_{ij},~<\psi|\psi>=1.
\label{orthonormal}
\end{equation}
First consider the special case
\begin{equation}
|\psi>=\sum_{i=1}^N \cos {\theta\over 2}|\psi_i>.
\label{psisp}
\end{equation}
 Then (\ref{orthonormal}) implies 
\begin{equation}
N\cos^2 {\theta\over 2}=1.
\label{N}
\end{equation}
It also follows from (\ref{distance}) that the state represented by 
$|\psi>$ has equal distance $\theta$ from all the states 
represented 
by 
$|\psi_i>$. Replacing $|\psi_i>$ by $e^{i\phi_i}|\psi_i>$ in 
(\ref{psisp}) 
or 
(\ref{psi}) does not change the distances between states. This is 
why 
$A_3$ was stated in terms of distances  in the state space.
It follows from $A_3$ that the probabilities of transition from 
$\psi$ 
to all $\psi_i$ are the same. Since these 
probabilities must add up to $1$, the probability of transition 
from 
$\psi$ to each $\psi_i$ is $1\over N$. But from  (\ref{N}) and 
(\ref{psisp})
\begin{equation}
{1\over N}=\cos^2 {\theta\over 2}=|<\psi|\psi_i>|^2
\label{1/N}
\end{equation}
which is therefore the probability of transition from 
$\psi$ to $\psi_i$.

Consider now the more general case of (\ref{psi}) in which the 
$c_i$s 
need not be equal. Introduce now an auxiliary system whose possible
states belong to an infinite dimensional Hilbert space. The state 
vector of the combined system is the direct product 
$|\psi>|\phi>$ , where the state $|\phi>$ of the auxiliary system does not 
interact with $|\psi>$ and 
both state vectors are normalized. The purpose of introducing the auxiliary 
system is to use the assumption $A_3$ on $|\psi>|\phi>$ in order to derive the 
the probabilities of transition from 
$\psi$ 
to all $\psi_i$. For each positive integer $n_i$, 
there exist orthonormal state vectors $\{|\phi_{ij}>, j=1,2,...,n_i\}$ 
such that  
\begin{equation}
|\phi> = \sum_{j=1}^{n_i}{1\over \sqrt{n_i}} |\phi_{ij}>
\label{auxiliary}
\end{equation}
Using (\ref{auxiliary}), the state vector of the combined system 
may be written as
\begin{equation}
|\psi>|\phi>= \sum_{i=1}^N \sum_{j=1}^{n_i}{c_i\over \sqrt{n_i}} 
|\psi_i>|\phi_{ij}>.
\label{state}
\end{equation}
Choose the positive integers $n_i$ so that for all $i,j$,
\begin{equation}
{c_i\over \sqrt{n_i}}\approx {c_j\over \sqrt{n_j}}.
\label{approx}
\end{equation}
If the ratios $c_i^2/c_k^2$ are rational for all $i$ and a fixed $k$, and $N$ is 
finite, then 
there exist positive integers
$n_i$s such that
$c_i^2=n_i/M$, where $ M$ is a common real denominator. Then
(\ref{approx}) is an exact
equality. If at least one of these ratios is irrational then, owing to 
the
rational numbers being dense in the real line, the 
positive integers $n_i$ may be made large enough to make 
(\ref{approx}) as close as possible to 
equalities. 

Then (\ref{state}) reduces to the special case considered before in 
which the 
coefficients are equal. Therefore, the probability of transition to 
$|\psi_i>|\phi_{ij}>$ is ${c_i^2\over {n_i}}$ ($j=1,2,...n_i$), using
(\ref{1/N}). Since there 
are $n_i$ such mutually exclusive 
states for a given $i$, the probability of transition to the state 
$|\psi_i>$ is $n_i\times {c_i^2\over {n_i}}=c_i^2=|<\psi|\psi_i>|^2$, 
using 
(\ref{psi}) and (\ref{orthonormal}). So the quantum 
probability rule has been derived from the assumptions 
$A_1$ to $A_4$. 
A somewhat similar argument has been given by Zurek (1998) using 
density matrices in the context of decoherence.

It follows 
that the probability of transition between Heisenberg 
states $|\alpha>$ and $|\beta>$ is 
$|<\alpha|\beta>|^2$ irrespective of the time elapsed between the 
observations of $|\alpha>$ and $|\beta>$. Hence, the converse of 
$A_3$, above, is also valid.

\section{Symmetries and Interactions}

If there are no laws then how do we explain the regularities 
we observe 
in the world? I shall argue here that the answer lies in 
symmetries. To make this argument, I shall first use the presently 
used laws of physics in order to obtain a characterization of all 
fundamental interactions in terms of symmetries. This statement 
will be so elegant and simple that it will suggest that we may 
discard the laws and keep the symmetries.

First consider the realization of quantum probabilities obtained in 
Section 5 in a simple experiment, namely diffraction 
through a narrow slit. In this phenomenon, particles that were 
initially prepared in the same state, after they go through the slit, 
end up at various points on a screen. It is well known that, the 
density of distribution of particles on the screen (the diffraction 
pattern), which gives the probability density for the possible final 
states in the limit of large number of particles, has an oscillatory behavior: 
It has a central maximum 
where the particles are most likely to strike, and the probability 
density decreases,  as would be expected, as we move away from 
this point. But then after reaching a minimum, strangely enough, 
the probability density increases, then decreases again to a 
minimum and increases again, and so on. This is very different 
from what would be expected classically, namely a monotonic 
decrease in probability density as we move away from the central point.

I believe that nature is telling us something very important about 
quantum geometry even in this simple experiment. It is that given 
two points $A$ and $B$ separated by a distance $\ell$, the 
translation operator
\begin{equation}
\exp(i{\bf p}\ell),
\label{translation}
\end{equation}
where $\bf p$ is the momentum (generator of translations) in the 
direction from $A$ to $B$, is more fundamental than the distance 
$\ell$ between $A$ and $B$, because the eigenvalues of (\ref{translation}) 
oscillate with $\ell$. It is (\ref{translation}) which relates 
appropriate pairs of quantum states (represented by vectors in 
Hilbert space), whereas $\ell$ may be regarded as a derived 
concept. Indeed, $\ell$ regarded as the  distance between $A$ and 
$B$ is an approximate concept because there is inevitably a 
fuzziness in the determination of points owing to the uncertainty 
principle, as shown for example by Wigner (1967, p. 62-69). But 
(\ref{translation}), 
regarded as a translation of any quantum state, has no such 
fuzziness. 

Now (\ref{translation}) is equivalent to the `modular momentum' 
introduced by Aharonov et al (1969) as a more fundamental 
variable than the momentum $\bf p$. They showed that this 
variable captures an essential non local aspect of quantum theory, 
which makes quantum physics fundamentally different from 
classical physics. To see this consider the wave packet
\begin{equation}
\psi_\alpha(x,y,z) ={1\over \sqrt{2}}f(x,y,z) + {1\over \sqrt{2}} 
\exp(i\alpha)f(x-\ell,y,z)
\end{equation}
where $f(x,y,z)$ is a normalized wave function such that $f(x,y,z) $ 
and $ f(x-\ell,y,z)$ are non overlapping functions. For example, 
$\psi_\alpha$ may be the superposition of two wave packets in a 
double slit interference experiment at the time when they are 
emerging from the two slits that are separated by a distance 
$\ell$ in the $x-$direction. Then
\begin{equation}
<\psi_\alpha|\exp(i{\bf p}\ell)|\psi_\alpha> = {1\over 2}\exp(i\alpha) .
\label{modular}
\end{equation}
So the expectation value of (\ref{translation}) gives the phase 
difference between the two superposed wave packets. What 
makes this non local is that no experiment performed on the wave 
packets at the given time could detect this phase difference 
between the non overlapping wave packets. But $\alpha$ may be 
observed when they interfere subsequently. This is unlike in 
classical physics where all experiments performed locally on the 
two wave packets can predict what will happen subsequently. In 
classical physics also the translational operator generated by  the 
momentum may be defined on the phase space, which is the classical analog of the 
Hilbert space, but it 
contains no more information than the momentum. But in 
quantum physics, (\ref{translation}) contains more information 
than $\bf p$, which is non local and observable.

The result (\ref{modular}) may be applied to the momentum 
distribution in the Aharonov-Bohm (AB) effect  on the electron 
beams interfering around a solenoid due to the magnetic field 
confined inside the solenoid.  Aharonov et al (1969) showed that 
although there is no exchange of momentum between each 
electron and the solenoid in the AB effect, for the obvious reason 
that there is no force on the electron due to the magnetic field, 
there is nevertheless a  non local exchange of modular momentum 
between them. The magnetic field causes a shift in the 
interference fringes due to the exchange of $\exp(i{\bf 
p}\ell)$. But the envelope of the interference pattern does not 
shift because there is no exchange of any of the moments of momentum ${\bf p}^n$ 
($n$ is any positive integer). Hence, 
(\ref{translation}) is a more fundamental observable than $\bf p$. 
Combining this with the earlier considerations in this section, I conclude that 
the group element $\exp(i{\bf p}\ell)$ is more fundamental 
than either $\bf p$ or $\ell$.

In the presence of the electromagnetic field,
$\exp(i{\bf p}\ell)$ is gauge dependent. The gauge invariant 
re-statement of the above is that in the AB effect the modular kinetic momentum 
$\exp i({\bf p}\ell- e\int_0^\ell A dx)$ is exchanged but not 
the kinetic momentum ${\bf p}- e A$. The last two operators are 
gauge invariant. Aharonov et al (1969) also showed that when 
there is a forceless interaction, there could be an 
exchange of modular energy in the absence of an exchange of 
energy. Also, the above considerations on the AB effect will also 
apply to the generalization of the AB effect to arbitrary gauge 
fields (see Wisnivesky and Aharonov (1967), Anandan(1970)). I 
therefore generalize the modular kinetic momentum covariantly, 
in an arbitrary gauge field, to the modular kinetic energy-
momentum
\begin{equation}
f_\gamma=P\exp \{i\int_0^{\ell^\mu} (p_\nu- A^k_\nu T_k)
dx^\nu\}
\label{gauge}
\end{equation}
where ${T_k}$ generate the gauge group, $A^k_\nu$ is the gauge 
potential and $P$ denotes path ordering.

To generalize this further to include the gravitational field, I shall 
use the principle of equivalence according to which at each space-
time point there exists a local Minkowski coordinate system $(x^a, 
a=0,1,2,3)$. The kinetic energy-momentum operator in this 
coordinate system is
$$\Pi_a\equiv p_a-  A^k_aT_k .$$
This may be written in an arbitrary coordinate system as
$$\theta_\mu^a \Pi_a \equiv \theta_\mu^a p_a-  A^k_\mu T_k ,$$
where $\theta_\mu^a$ is the canonical 1-form (Kobayashi and 
Nomizu (1963) or the solder form (Trautman, 1973, Anandan 
,1993b), so called because it solders the 1-forms in the local 
inertial frame to the cotangent space of space-time at each point. The 
generalization of (\ref{gauge}) in the 
presence of gravity and gauges fields is then

\begin{equation}
g_\gamma = Pexp(- i\int_{\gamma} \Gamma_\mu dx^\mu 
),
\label{symmetry}
\end{equation}
where
\begin{equation}
\Gamma_\mu ={\theta_\mu}^a 
P_a +{1\over2}{{{\omega}_\mu}^a}_b {M^b}_a -A^k_\mu T_k .
\label{connection}
\end{equation}
Here, $P_a$ generate space-time translations, ${M^b}_a$ generate 
Lorentz transformations and $T_k$ generate the gauge group $G$. 
The $\theta$, $\omega$ and $A$ are defined along the curve 
$\gamma$. ($\theta$ and $\omega$ should not be thought of as 
functions of space in the definiton of $g_\gamma$ because this 
would make $g_\gamma$ non unitary.)
Hence,
$g_\gamma ~\epsilon~ $ Poincare group $\times ~G$.

$\gamma$ may be any open or closed curve. When $\gamma$ is an infinitesimal 
closed curve spanning an area element $d\sigma^{\mu\nu}$, using (\ref{symmetry}) 
and the Lie algebra relations of the Poincare group and the gauge group, we 
obtain (Anandan, 1980, 1996)
\begin{equation}
g_\gamma =1+i({Q_{\mu\nu}}^aP_a +{1\over2}{{R_{\mu\nu}}^a}_b {M^b}_a 
+{F_{\mu\nu}}^kT_k){ d\sigma^{\mu\nu}\over 2}
\end{equation}
where the $2-$forms
\begin{equation}
Q^a=d\theta^a +{{\omega}^a}_b\wedge\theta^b ,{R^a}_b 
=d{\omega^a}_b+{\omega^a}_c\wedge{\omega^c}_b, F=dA^k + C^k_{mn}A^m\wedge A^n
\end{equation}
are called the torsion, linear curvature, and the Yang-Mills field strength, 
respectively, and  $ C^k_{mn}$ are the structure constants of the gauge Lie 
algebra.
Whenever two quantum systems interact via 
gravity and gauge fields, it can be shown that there exist curves $\gamma$ so 
that 
$g_\gamma$ is exchanged between the two systems.

This result was obtained from laws which had the above symmetries, 
at 
least locally. But since $g_\gamma $ is what is observed 
\footnote{In practice, we observe Hermitian operators such as 
$g_{\gamma} +{g_\gamma}^\dagger$ and  $i(g_\gamma 
-{g_\gamma}^\dagger)$ from which of course $g_\gamma$ 
may be determined. See Anandan (1986a,b) for a discussion of 
how 
these Hermitian operators may be observed in interference 
experiments.} , it may 
be possible to keep it and discard the laws. 
Then symmetries need not be regarded as originating as invariances of 
laws but instead regarded as relations between 
states. 
These relations, and hence the symmetries, are universal in that 
they have representations in every Hilbert space. (The gauge 
group part of the symmetries would be trivial in representations 
in which the gauge charges are zero.)
They may therefore be regarded as defining the geometry of 
quantum theory.

The above considerations suggest the following principle which 
characterizes all interactions: {\it Two quantum systems interact if 
and only if there is an exchange of a symmetry group element 
between them.} In terms of the usual quantum theory, this means that the 
expectation value of the symmetry group element for each of the states of the 
quantum system changes, but not for the state of the combined system. If the 
symmetry group $S$ is chosen to be that of 
the standard model, i.e.  $S= P\times G$, where $P$ is the 
Fermionic covering group of the Poincare group (semi-direct 
product of $SL(2,C)$ with the space-time translational group) and 
$G=U(1)\times SU(2)\times SU(3)$, then I conjecture that in an appropriate limit 
the usual 
gravity and gauge fields are obtained. The appropriate limit means that the 
systems which produce the gravity and gauge fields may be treated classically. 
In this limit the group element which is exchanged may be written in the form 
(\ref{symmetry}) so that classical gravity and gauge fields may be obtained from 
(\ref{connection}). But we may need to change $S$ when the 
standard model is superseded by a deeper theory.  The above 
principle would then still hold and give us a new set of fields corresponding to 
the new symmetry group. I shall therefore call these fields, corresponding to 
arbitrary symmetry groups, symmetry fields.

In the above treatment, gravitational and gauge fields were 
treated classically. When the fields are quantized, the probability 
amplitude for a process is at present obtained by adding integrals 
represented by Feynman diagrams. At each vertex of a Feynman 
diagram the energy-momentum is exchanged. But it may be useful 
to reformulate quantum field theory in terms of exchange of 
modular energy-momentum. At distance scales of the order of
Planck length, space-time geometry breaks down, and the usual 
Feynman diagrams may not be meaningful. Also, in 
(\ref{symmetry}), $\gamma$ cannot then be meaningfully 
defined. However, we may then be able to replace 
(\ref{symmetry}) by an element of the symmetry group that does 
not require a space-time curve $\gamma$ for its definition. Such a theory would be a 
quantum theory of gravity. Since quantum gravity is expected to unify all the 
interactions, the unified treatment of all the fundamental interactions in 
(\ref{symmetry}) suggests that it may be useful in constructing quantum gravity.

An advantage of the present approach is that non local interactions, such as the 
non local exchange of modular momentum between the solenoid and the charge in 
the AB effect and its generalization to gauge fields and gravitation, is easily 
treated. But in the AB effect for example, the change in the moduluar momentum 
undergone by the charged wave function gives the AB phase shift around a closed 
curve because of the $U$-process undergone by the interfering beams.
Also, the $U$-process due to Schr\"odinger  evolution undergone by the 
state vector in the Schr\"odinger picture or the observables in the 
Heisenberg picture was associated earlier with ensembles of 
identical 
quantum systems because this determines the probabilities of processes of 
individual systems. The Hamiltonian that generates the $U$ process 
comes from a Lagrangian. It may appear that the specification of the Lagrangian 
constitutes a law. But in fact, the general form of the Lagrangian for a system 
of non relativistic interacting particles may be obtained from Galilean 
symmetries (Landau and Lifshitz, 1976; Lawrie, 1994).  
In relativistic quantum field theory, the fields in the Lagrangian must 
provide 
representations of the symmetry group and the Lagrangian is 
invariant under this group. This is a major constraint on the fields 
and
the Lagrangian.

But even after the fields are specified, 
there are still an infinite number of Lagrangians that are invariant 
under the symmetry group. However, the requirement of 
renormalizability 
places a severe restriction on the possible Lagrangians. 
In the case of 
the symmetry group $S$ of the standard model, after specifying the fields and 
the 
coupling 
constants (which are the contingencies of 
this 
model), the requirement of renormalizability pretty much 
uniquely 
determines the Lagrangian of all the fields, excluding gravity, as mentioned by 
Weinberg (1980). A quantum theoretic description of the gravitational field 
cannot be obtained this way at present, which I believe to be due to an 
insufficient understanding at present of the connection between symmetries and 
the gravitational field. One of the purposes of the present article is to 
elucidate this connection, which hopefully would help in constructing a quantum 
theory of gravity. 
In 
a future theory, renormalizability may be realized as being due to 
logical self-consistency of the theory.

I therefore make the following conjecture: Regularities in physical 
phenomena are due to 
symmetries, logical or mathematical consistency, and 
contingencies. The contingencies here include algortithmically 
undecidable propositions which exist, according to G\"odel's 
theorem, in any axiomatic system that is at least as rich as the 
natural numbers. In view of G\"odel's theorem, it may 
not be possible to construct a physical theory that is due entirely 
to logical or mathematical necessity. Also, at present it seems possible for 
the world to have, in principle, symmetries that are different from the ones we 
observe, or have no symmetries at all, without violating logical or 
mathematical necessity. So, if we choose a particular symmetry group, in the new 
paradigm of symmetries proposed here, it would restrict the physical theory to a 
particular, but it seems contingent, class of mathematical models.

\section{Quantum Geometry}

There are parallels in
physical geometry to the two views of having 
dynamical laws, or not having dynamical laws but having 
symmetries. In the former case, 
it is 
natural to adopt the Riemannian geometry of space-time because 
the 
dynamical laws give evolutions which are best described 
in space-time. 
E.g. geodesics are world-lines of free particles in classical general 
relativity. 
But it was pointed out that a different conception of geometry is 
better suited for quantum theory (Anandan, 1980a).  This is based on 
Klein's 
Erlanger program, according to which a geometry is a set of 
properties invariant under a group of transformations acting on a 
set 
of points. It is difficult to take this set to be space-time, because 
space-time points are not observable owing to the uncertainty 
principle in quantum theory, as already mentioned. Also, it does not seem
plausible to 
talk of an electron as being immersed in space-time, because its 
states belong to a Hilbert space, which has a very different 
geometry 
and interpretation.

It therefore seems reasonable to generalize Klein's Erlanger 
program so that the symmetry group does not act on a universal 
set 
of points. Instead, I shall let the same symmetry group $S$ to 
simply act 
on each Hilbert space, which is the set of possible states of a 
system.
Each $g\epsilon S$ is universal in the sense that the relation it
determines between states $\alpha$ and $\beta \equiv g\alpha$ 
is
independent of $\alpha$, including which Hilbert space $\alpha$ 
belongs
to. Also the relation defined by the
condition that a pair of states are related by arbitrary
$g\epsilon S$ is an equivalence relation.  The last two properties
justify associating a geometry directly with $S$.
The available evidence at present is consistent with 
$S\equiv$  Poincare group $\times ~G$, where
$G=  U(1)\times SU(2)\times SU(3)$. This universal symmetry
group $S$, as shown in the previous section, gives rise to all the fundamental 
interactions.
Replacing dynamical laws by symmetries as the fundamental 
concept in physics 
corresponds 
to changing the physical geometry to the above concept of 
geometry 
in which 
symmetries directly relate equivalent states without being 
required
to act on a universal set of
points, such as the Riemannian space-time.

The Riemannian space-time enabled the natural inclusion of the
classical gravitational field as the metrical relations between 
space-time
points.
Similarly, the generalized Klein geometry described above enables 
the
natural inclusion of all fundamental interactions as symmetry  
relations
between
quantum states. This is provided by (\ref{symmetry}) in an 
approximate 
manner
because $\gamma$ is a space-time curve, which has no operational meaning in 
quantum physics. A more precise 
treatment may give
the quantized fields instead of the classical fields contained in
(\ref{symmetry}).

The acceptance of the primary role of symmetries in physics would remove the 
mystery of why there are complex numbers in quantum mechanics. This is because 
the representations of groups act more naturally on complex vector spaces than 
on real vector spaces. For example, the $U(1)$ group has a real faithful 
representation 
$SO(2, {\cal R})$, where $\cal R$ is the real line. But the matrices of this 
representation and its Lie algebra do 
not have eigenvalues or eigenvectors, if we restrict to the field of real 
numbers. The elements of $SO(2,{\cal R})$ may be written in the form $\exp(\phi 
X)$, where $\phi\epsilon {\cal R}$ and $X^2=-I$. Hence, $X$ defines a complex 
structure on the vector space
${\cal R}^2$ on which $SO(2,{\cal R})$ acts. Since $X$ commutes with $SO(2,{\cal 
R})$, this complex structure is invariant under $SO(2,{\cal R})$. It is then 
natural to treat this vector space as a one-dimensional complex vector space 
$\cal C$, with $X$ replaced by $i$ and $SO(2,{\cal R})$ correspondingly replaced 
by $U(1)$. 
The different representations of the $U(1)$ group then correspond physically to 
different charges. The compactness of the $U(1)$ group then implies that charges 
are integer multiples of a fundamental charge (see for example Yang, 1969), 
which is consistent with 
observation. 

In the case of the Lorentz group, the existence of Fermions implies that it 
should be $SL(2,{\cal C})$ and not its $(2-1)$ isomorphic group $SO(3,1,{\cal 
R})$. The fundamental representation of $SL(2,{\cal C})$ acts on ${\cal C}^2$, 
which is the space of spinors. Here also we may regard this action as the action 
of a real linear group on ${\cal R}^4$, which however contains a complex 
structure $Y$ ($Y^2=-I$) that commutes with this group. So, it is natural to 
regard this ${\cal R}^4$ as ${\cal C}^2$. The action of $Y$ on this ${\cal C}^2$ 
is simply $iI$, where $I$ is the identity transformation. But a vector in any 
tangent space $V$ of space-time is a quadratic combination of the spinors and 
their complex conjugates (or the vectors in the above ${\cal R}^4$) at that 
point. Therefore, the action of $Y$ corresponds to the identity operation on 
$V$. This explains why the complex structure $Y$ was `hidden' prior to the 
discovery of Fermions. It also explains why general relativity, which was 
formulated in terms of tensors that take their values in the tensor products of 
$V$ with itself and its dual, does not need complex numbers. But it should be 
noted that even prior to the discovery of Fermions, complex numbers were used in 
quantum mechanics because of unitary representations of symmetry groups. As 
remarked above, these representations are more natural than real representations 
because of the existence of eigenvalues and eigenvectors over the field $\cal 
C$.

The view that there is no law corresponds to our experience of a 
`flow of
time', which is a mystery in the paradigm of laws. As mentioned 
above, in
the latter paradigm space-time geometry is the appropriate 
geometry.
But if everything is laid out on space-time, there does not seem to 
be any
room for time to `flow'. Moreover, the time of our consciousness has 
an
arrow due to our remembering our past, which is unique, but we 
do not know
of our uncertain future. But this is the same as the 
arrow of
time determined by pheonomena, obtained in section 3, because the final states 
of a 
phenomenon, as
expressed by $A_1$ above, is uncertain in general. So, the present approach has 
the advantage that it is in accordance with our conscious experience of time, 
which may be the most immediate experience we have, and which therefore science 
should take into account.

\section{Comparison With Other Approaches To Quantum 
Theory}

The present approach will now be compared with some other 
interpretations of quantum theory. Two consistent interpretations 
of 
quantum theory within the paradigm of laws are the Bohm 
(1952) (see also Holland, 1993), and the Everett (1957)
interpretations. 
Both of them take Schrodinger's equation seriously as the 
deterministic law governing quantum  evolution.  The Bohm 
interpretation avoids the `many worlds' of Everett by postulating 
a 
dual ontology that requires the existence of particles as well as the 
wave. This also has the advantage that it is possible to introduce 
probabilities which are consistent with experiment by associating 
them with the particles while the wave undergoes deterministic 
evolution. Whereas in the Everett interpretation, since there is 
only a 
wave that undergoes deterministic evolution, it is not possible to 
explain the probabilities observed in quantum experiments by 
simply postulating a measure on the relative states (Everett, 1957). 
In 
the Bohm interpretation, on the other hand, it is not clear why the 
wave function which deterministically guides the particle should 
also 
give the probability density for finding it.

The present approach does not require a dual ontology as in the 
Bohm interpretation. It resembles most the Copenhagen 
interpretation, but differs from it in not making an arbitrary 
division 
between the classical and quantum worlds. By including protective 
measurements, which were unknown to the founders of the 
Copenhagen interpretation, the present approach enables an 
extended wave function to be treated as real, instead of treating 
only 
the localized `events' associated with the `classical' measuring 
apparatus
as real as 
in the Copenhagen interpretation. The indeterminacy of quantum theory was 
introduced here by the denial of deterministic evolutionary laws governing the 
state, and not by denying the reality of the wave function and treating it as 
containing a catalog of probabilities as done by the Copenhagen interpretation. 
The present interpretation shares, however, with the 
Copenhagen and other interpretations the mystery of what a 
measurement really is. 

There are two advantages to the present approach over the other 
interpretations of quantum theory: 1) The indeterminacy of 
quantum 
mechanics is built in right from the beginning by denying the 
existence 
of deterministic laws. In the other interpretations, which work 
within the paradigm of laws, it is not clear why the dynamical 
laws which are
believed to exist are not deterministic;
this intrinsic indeterminacy is a 
mystery and needs to be postulated in an arbitrary 
manner\footnote
{The Copenhagen interpretation tries to rationalize this 
indeterminacy 
within the pardigm of laws by saying that the observation of a 
microscopic system disturbs it, as in the well known Heisenberg 
microscope experiment. The use of the wave function to describe 
the 
state of the microscopic system is then argued as being necessary 
in order to represent this indeterminacy. But this argument cannot 
be made in the case of 
protective measurements of the quantum system 
which do not disturb the state of the system
represented by its wave function, which is directly observed.}. 

2) If there are also no probabilistic laws but only 
symmetries at a fundamental level, then
the probabilities
are constrained only by the symmetries. Hence, the probabilities
must depend on the invariants under these symmetries 
which include the geometry of the Hilbert space.
So it is not surprising that in the derivation
of probabilities, in section 5, the
geometry of the Hilbert space played a role. This throws light on 
another long standing mystery of quantum mechanics, namely 
why 
the probabilities come from the {\it geometry} 
of the Hilbert space\footnote{
For a different geometric approach to the stochasticity of quantum 
mechanics see Hughston (1996).}, unlike in general relativity 
where 
geometry plays no stochastic role. 

\section{Further Discussion and Conclusion}

In the foregoing considerations, the paradigm 
of laws was rejected because it is based on the metaphysical assumption that 
physical systems should in some mysterious way be `railroaded' into obeying 
dynamical laws. Such a `railroad' cannot be observed. For example, the evolution 
of the state of an electron between successive observations, say at the source 
and the detector, cannot be operationally defined. In the Bohm interpretation, 
considered in the previous section, the electron may be assigned a trajectory 
between the source and the detector, but this trajectory is not observable. To 
describe the evolution of the wave function of the electron as it interacts with 
other macroscopic systems requires the introduction of the `many worlds' of 
Everett, which are also not observable.
On the other hand, the generators of the symmetry group are conserved 
quantities, such as charge or total momentum, which are observable. The 
existence of these conserved quantities is also shown from the fact that they 
generate gauge fields and gravity which have observable effects. 

The history of physics has shown the 
usefulness of rejecting metaphysical assumptions, e.g. Einstein's 
rejection of absolute simultaneity in developing special relativity 
and the notion of a global inertial frame in developing general 
relativity.
The rejection of the metaphysical assumption of laws led 
immediately to the intrinsic indeterminacy of quantum theory, in 
section 2, which in turn gave an arrow of time, in section 3. In section 5, the 
observed probabilities of quantum theory were also obtained in a 
natural manner. The postulate that symmetries should play the 
fundamental role in the new proposed paradigm naturally gave, in 
section 6, the observed gravitational and gauge fields for the symmetries of the 
standard model.

The last result may also be used to argue for the reality of symmetries. 
(\ref{symmetry}), in section 6, may be regarded as an alteration of the 
symmetries experienced by a quantum state due to gravity and 
gauge fields. This changes the probabilities of processes undergone 
by the quantum system that experiences these fields. The 
quantum system may then be expected to  react
back to
modify the fields contained in (\ref{symmetry}). We know this to 
be the case experimentally: all systems modify the gravitational 
fields, charged systems modify the electromagnetic field etc. So, if 
the system is affected by the fields, which depends on the 
particular representation of the symmetry group to which it 
belongs to, then it reacts back on the fields. A criterion for reality 
has
been proposed by Anandan and Brown(1995) according to which 
an object may be 
regarded
as real if it satisfies the action-reaction principle. According to this
criterion, the symmetry group element (\ref{symmetry}), which replaces the usual 
gravitational and gauge fields, may be 
regarded as
real. I believe that it is these symmetry group elements (\ref{symmetry}), 
instead of the usual fields, which should be used to construct a quantum theory 
of gravity.

We may draw encouragement for the last remark from the following historical 
fact. As pointed out by Yang (1987), the importance of the 
electromagnetic phase factor,
\begin{equation}
 exp(- {ie\over\hbar}\int_{\gamma} A_\mu dx^\mu 
),
\label{abphase}
\end{equation}
was recognized by Schr\"odinger (1922), in his study of Weyl's 
gauge theory, four years before he introduced the wave function. I 
wish to emphasize that (\ref{abphase}) is an element of the $U(1)$ 
symmetry group, and is part of (\ref{symmetry}), which belongs 
to the entire symmetry group. From the present point of view, 
(\ref{abphase}) is not only historically but also logically prior to the 
Schr\"odinger equation introduced in 1926.  Similarly, 
(\ref{symmetry}) may be a precursor to a fully developed 
quantum theory of gravity which is yet to come. 

If the last conjecture is realized then there may still be the 
ultimate 
question of why (\ref{symmetry}) belongs to a particular 
symmetry group $S$
and not some other symmetry group. The $U(1)$, $SU(2)$ and 
$SU(3)$ 
symmetries, which are contained in $S$ according to the standard model, 
are the simplest unitary 
symmetries and may be justified on the grounds of simplicity. But 
why should $S$ contain the Poincare group of symmetries instead 
of 
some other symmetries? The anthropic principle may 
provide 
the answer to this question\footnote
{I thank Dennis Sciama for a discussion concerning this point.}. 
According to this principle, as 
interpreted here,
there are parallel universes in which there could be different 
symmetries, and hence different effective laws that are obtained 
from these symmetries, logical self-consistency, and contingencies.  The 
universe 
in 
which we live must be one in which the symmetries are such that 
they allow life to evolve. This 
places a restriction on the symmetries in our universe.  However, 
we should also be open to the possibility that quantum gravity 
may modify $S$ to some other symmetry group.

The purpose of physics is to simplify our understanding of nature. 
Simplification is often accompanied by immense progress, not only in physics but 
also in all of science. Giving up trying to explain the initial conditions in 
favor of laws as our basis for understanding nature, mentioned at the beginning 
of this article, led to an enormous simplification. This also resulted in 
tremendous progress in physics. Similarly, foregoing laws and accepting 
symmetries as our basis for understanding nature would lead to another enormous 
simplification, and perhaps also much more progress. This is particularly 
relevant to the construction of quantum gravity. Instead of trying to discover 
new laws, we should perhaps focus on understanding how symmetries give rise to 
all the fundamental interactions in nature, including gravity. The quantum 
mechanical nature of these interactions, as discussed above, is then a natural 
consequence of the denial of the existence of fundamental laws, the central role 
of symmetries, and the nature of the states of the systems that undergo these 
interactions.
 
\bigskip

\noindent{ACKNOWLEDGMENTS}
\bigskip

I thank Yakir Aharonov for useful discussions, especially 
concerning modular variables.  I also thank Roger Penrose,
W. H. Zurek  and Ralph Howard for
stimulating discussions, and Philip Pearle for useful comments 
on an 
earlier version of this paper. The F. C. Donders Chair the University 
of Utrecht, which provided me support  when part of this work 
was being done, is gratefully acknowledged. This research was 
also partially 
supported by the NSF grant no. 9601280.
\newpage
\noindent{REFERENCES}
\bigskip

\noindent
Aharonov Y., Pendleton H. and Peterson A. 1969 Int. J. Theoretical 
Phys., vol. 2, 213-230.

\noindent
Aharonov Y. , Anandan J., and Vaidman L. 1993, Phys. Rev. A
{\bf
47,} 4616; Aharonov Y.  and Vaidman L., Phys. Lett. A 1993 {\bf
178,} 38.

\noindent
Anandan J. 1979 Nuov. Cim. {\bf 53A,} 221.

\noindent
Anandan J. 1980a, Foundations of Physics {\bf 10,} 601-629.

\noindent
Anandan J. 1980b, in {\it Quantum Theory and Gravitation}, 
edited by A. R. Marlow (Academic Press, New York ), p. 157. 

\noindent
Anandan J. 1986a Phys. Rev. D {\bf 33,} 2280-2287.

\noindent
Anandan J. 1986b, in
Proceedings of the Erice school on "Topological Properties and 
Global Structure of Space-time", May 12-22, 1985, edited by P.G. 
Bergmann (Plenum, 1986).

\noindent
Anandan J. 1991, Foundations of Physics {\bf 21,} No. 11, 
1265-1284.

\noindent
Anandan J. 1993a, Foundations of Physics
Letters {\bf 6,} 503.

\noindent
Anandan J. 1993b, in {\it Directions in General Relativity}, eds. B.L. Hu, M.P. 
Ryan and C. V. Vishveshwara (Cambridge University Press, Cambridge), vol. 1, p 
10-20.

\noindent
Anandan J. 1996, Phys. Rev. D  {\bf 53,} No. 2, 779-786, gr-
qc/9507049.

\noindent
Anandan J. and Aharonov Y. 1990, Physical Review Letters {\bf 
65,} 
1697-1700.

\noindent
Anandan J.  and Brown H. R. 1995, Found. Phys. {\bf 25,} 349-360.

\noindent
Bohm D. 1952, Phys. Rev., 
{\bf 85,} 166-193.

\noindent
Diosi L., Phys. Rev. A {\bf 40,} 1165-1174 
(1989).

\noindent
Everett H. 1957, Rev. Mod. Phys.  {\bf 29,} 454-462.

\noindent
Gnedenko B. V. 1967, {\it The Theory of Proabibility} (Chelsea, 
NY), 
ch. 1, sec. 4.

\noindent
Ghiradi G. C., Rimini  A.,  and Weber T. 1986, Phys. Rev. D, {\bf 34,} 
470-491.

\noindent
Ghiradi G. C., Grassi R. and Rimini A. 1990, Phys. Rev. D, 
{\bf 42,} 1057-1064.

\noindent
Holland P. R. , 1993 {\it The Quantum Theory of Motion} 
(Cambridge 
Univ. Press, Cambridge).

\noindent
Hughston L. P. 1996, Proc. Roy. Soc. London A {\bf 452,} 
953-979.

\noindent
Klein, Felix 1872, A Comparative Review of Recent Reserches in
Geometry (Erlanger) [English translation in N.Y. Math. Soc. 
Bull. 2
(July), 215 (1893)]. 

\noindent
Kobayashi S.  and Nomizu K. 1963, {\it Foundations of Differential 
Geometry} (John Wiley, New York) p. 118.

\noindent
Landau L. D. and Lifshitz E. M. 1976, {\it Mechanics} (Pergamon Press, Oxford), 
$\S$ 3, $\S$ 4.

\noindent
Lawrie Ian D. 1994, {\it A Unified Grand Tour of Theoretical Physics} (IOP 
Publishing, Bristol and Philadelphia), p. 51, 52.

\noindent
Pearle P. 1986, in{\it Quantum Concepts in Space and Time,} eds. 
R. Penrose 
and C. J. Isham (Clarendon Press, ), 84-108. 

\noindent
Pearle P. 1989, Phys. Rev. 
A {\bf 39,} 2277-2289.

\noindent
Peirce, C. S. 1891, `The Architecture of Theories'' in the Monist, reprinted in 
{\it Philosophical Writings of Peirce}, edited by J. Buchler (Dover, NY, 1955).

\noindent
Penrose, Roger 1989, {\it The Emperor's New Mind} (Oxford Univ. 
Press, 
Oxford.

\noindent
Penrose R. 1996, Gen. Rel. and Grav., 
{\bf28,} 581-600.

\noindent
Schr\"odinger E. 1922, Z. f. Phys. 12, 13.

\noindent
Smolin, Lee 1997, {\it The Life of the Cosmos} (Oxford Univ. Press, Oxford).

\noindent
Trautman A. 1973, in {\it The Physicist's Conception of Nature}, 
edited 
by J. Mehra (Reidel, Holland).

\noindent
van Fraassen, Bas c. 1989, {\it Laws and Symmetry} (Oxford Univ. Press).

\noindent
Weinberg S. 1980, Rev. of Mod. Phys. {\bf 52,} 515-523. 

\noindent
Wheeler, John A. 1980, in {\it Some Strangeness in the Proportion}, edited by 
Harry Woolf (Addison-Wesley, Reading, MA) P. 341-375.

\noindent
Wheeler, John A. 1984, in Caianello Celbration volume edited by A. Giovannini, 
M. Marinaro, F. Mancini and A. Rimini. 

\noindent
Wheeler, John A. 1990, in {\it Proceedings of the Third International Symposium: 
Foundations of Quantum Mechanics in the Light of New Technology}, edited by S. 
Kobayashi, H. Ezawa, Y. Murayama, and S. Nomura (The Physical Society of Japan, 
Tokyo)

\noindent
Wheeler, John A. 1994, in {\it Quantum Coherence and Reality}, edited by J. S. 
Anandan and J. L. Safko (World Scientific, Singapore), p. 281-297.

\noindent
Wigner, Eugene 1967, {\it Symmetries and Reflections} 
(Indiana U. P., Bloomington/London).

\noindent
Wisnivesky D. and Aharonov Y. 1967,
Ann, of Phys. {\bf 45,} 479.

\noindent
Yang C. N. 1969, Phys. Rev. D {\bf 1,} 2360.

\noindent
Yang C. N. 1987, in {\it Schrödinger: Centenary Celebration of a 
Polymath}, edited by C. W. Kilmster (CUP, Cambridge).

\noindent
Zurek W. H. 1998, Phil. Trans. R. Soc. Lond. A {\bf 356,} 1793-
1821.

\end{document}